\documentclass[preprint]{aastex}

\shorttitle{}
\shortauthors{Fang et al.}

\begin{document}

\title{Modelling of Reflective Propagating Slow-mode Wave in a Flaring Loop}
\author{X.~Fang, D.~Yuan, T.~Van Doorsselaere, R.~Keppens, and C.~Xia}
\affil{Centre for mathematical Plasma Astrophysics, Department of Mathematics, KU Leuven, Celestijnenlaan 200B, 3001 Leuven, Belgium}

\begin{abstract}
Quasi-periodic propagating intensity disturbances have been observed in large coronal loops in EUV images over a decade, and are widely accepted to be slow magnetosonic waves. However, spectroscopic observations from Hinode/EIS revealed their association with persistent coronal upflows, making this interpretation debatable. We perform a 2.5D magnetohydrodynamic simulation to imitate the chromospheric evaporation and the following reflected patterns in a flare loop. Our model encompasses the corona, transition region, and chromosphere. We demonstrate that the quasi periodic propagating intensity variations captured by the synthesized \textit{Solar Dynamics Observatory}/Atmospheric Imaging Assembly (AIA) 131, 94~\AA~emission images match the previous observations well. With particle tracers in the simulation, we confirm that these quasi periodic propagating intensity variations consist of reflected slow mode waves and mass flows with an average speed of 310 km/s in an 80 Mm length loop with an average temperature of 9 MK. With the synthesized Doppler shift velocity and intensity maps of the \textit{Solar and Heliospheric Observatory}/Solar Ultraviolet Measurement of Emitted Radiation (SUMER) Fe XIX line emission, we confirm that these reflected slow mode waves are propagating waves. 
\end{abstract}


\section{Introduction}
The study of magnetohydrodynamic (MHD) waves in the solar atmosphere is an independent tool to understand the energy release processes, particle acceleration or heating mechanisms and to diagnose the plasma parameters indirectly by coronal seismology \citep{roberts00,moortel12,liu14}. MHD seismology was successfully applied in estimating the coronal magnetic field \citep{nakariakov01}, transverse loop structuring \citep{goossens02,aschwanden03}, polytropic index and thermal conduction
coefficient \citep{doorsselaere11b}, and the magnetic topology of sunspots \citep{yuan14a,yuan14b}. 

Standing longitudinal slow-mode oscillations were first discovered in the Doppler shift of hot emission lines (i.e., \ion{Fe}{19} and \ion{Fe}{21}) with formation temperature greater than 6 MK, by the Solar Ultraviolet Measurements of Emitted Radiation (SUMER) spectrograph onboard the Solar and Heliospheric Observatory (SOHO) \citep{wang02,wang03a,wang03b,wang11}. Similar Doppler-shift oscillations have been detected by Yohkoh/BCS in even hotter emission lines of S XV and Ca XIX, with formation temperature 12$\sim$14 MK \citep{mariska05,mariska06}. The oscillations are strongly damped within a couple of periods and are usually observed in association with the soft X-ray brightenings or even up to M-class flares \citep{wang07}.

Excitation of slow magnetoacoustic oscillations in hot coronal loops has been intensively studied theoretically. The compressible nature of the longitudinal oscillations and their long periods led to their interpretation in terms of standing slow magnetoacoustic oscillations damped due to thermal conduction \citep{ofman02}. In order to explain the observed damping time of the oscillations and demonstrate the robustness of this interpretation, several authors included other physical effects   \citep{nakariakov04,tsiklauri04,taroyan05,selwa05,selwa07,taroyan07,gruszecki11}, accounting for viscosity, multi-dimensional geometry, stratification, nonlinear steepening, and mode coupling. 

Quasi-periodic pulsations (QPP) observed in solar and stellar flares have been intensively studied for several decades \citep{nakariakov09,anfinogentov13}. The origin of QPPs still remains unclear, but one of the widely accepted theories is the modulation of QPP by magnetohydrodynamic (MHD) oscillations. Short period (sub-minute) oscillations are believed to be induced by fast mode waves, while those with periods of tens of seconds are ascribed to modulations by slow mode MHD waves \citep[e.g.][]{doorsselaere11a}.  Coronal MHD oscillations are  directly seen in various bands with modern instruments with high temporal and spatial resolution, which provide researchers with MHD wave diagnostics to identify physical conditions in flaring sites and mechanisms operating in them. 

Recently, the Solar Dynamics Observatory (SDO)/Atmospheric Imaging Assembly (AIA) provided high temporal and spatial resolution observations of slow mode oscillations in the solar corona. The first simultaneous observations of the electron density and EUV intensity oscillations were reported by  \citet{kim12} through Nobeyama 17 GHz and AIA 335\AA~channels, respectively. Kumar et al. (2013) reported the first direct observation of a propagating EUV disturbance (i.e., slow mode wave) in hot coronal arcade loops captured only in the AIA 131 and 94 channels. The wave was excited by an impulsive flare which occurred at one of the footpoints of the arcade loops. It showed multiple reflections between the opposite footpoints of the arcade loops (Kumar et al. 2013, 2015).
 The observed properties of these oscillations match the SUMER Doppler-shift oscillations associated with the slow magnetoacoustic mode. However, \citet{wang11} interprets the SUMER Doppler-shift oscillations as standing slow waves due to the associated intensity variations, which show roughly a quarter-period phase delay to the Doppler signal in some cases.

In this paper, we investigate the slow magnetoacoustic oscillations in a flare loop by a 2.5D magnetohydrodynamic simulation. The paper is then organized as follows: in \textsection2 we describe the numerical setup; in \textsection3 we show results of simulations and discuss the details; and conclusions are drawn in \textsection4.

\section{Computational Aspects}\label{setup}

\subsection{Governing Equations and Initial Setup}
Our numerical setup includes gravity, anisotropic thermal conduction and radiative cooling and parametrized heating terms, in a domain of -40 Mm $\leq  x \leq 40$ Mm and 0 $\leq  y \leq 50$ Mm.
The governing equations are as follows:
\begin{equation}
 \frac{\partial{\rho}}{\partial{t}}+\nabla\cdot\left(\rho \textbf{v} \right)= 0,
\end{equation}
\begin{equation}
 \frac{\partial{\left(\rho\textbf{v}\right)}}{\partial{t}}+\nabla\cdot\left(
\rho\textbf{vv}+p_{tot}\textbf{I}-\frac{\textbf{BB}}{\mu_{0}} \right)=\rho \textbf{g},
\end{equation}
\begin{equation}
 \frac{\partial{E}}{\partial{t}}+\nabla\cdot\left(E\textbf{v}+p_{tot}\textbf{v}-
\frac{\textbf{v}\cdot\textbf{B}}{\mu_{0}}\textbf{B}\right)=\rho\textbf{g}\cdot\textbf{v}+
\nabla\cdot\left( \vec{\kappa}\cdot\nabla T\right)-Q+H,
\end{equation}
\begin{equation}
 \frac{\partial{\textbf{B}}}{\partial{t}}+\nabla\cdot\left(\textbf{vB}-
\textbf{Bv}\right)=0,
\end{equation} 
where $T, \rho, \textbf{B}, \textbf{v}$, and $\textbf{I}$ are respectively temperature, density, magnetic field, velocity, and unit tensor. The total energy density is $E=p/\left(\gamma-1\right)+\rho v^{2}/2+B^{2}/2\mu_{0}$ and the total pressure is $p_{tot}\equiv p+B^{2}/2\mu_{0}$; $\textbf{g}=g_{0}R^{2}_{\odot}/\left(R_{\odot}+y\right)^{2}$\textbf{$\hat{y}$} is the solar surface gravitational acceleration with $g_{0}$ as $-274\,\mathrm{m}/\mathrm{s}^2$; $H$ and $Q$ are respectively the heating and radiative loss terms; $\vec{\kappa}$ is the thermal conductivity tensor. Assuming a 10:1 abundance of hydrogen and helium of completely ionized plasma, we obtain $\rho=1.4m_{\rm{p}}n_{\rm{H}}$, where $m_{\rm{p}}$ is the proton mass and $n_{\rm{H}}$ is the number density of hydrogen. We use the ideal gas law $p=2.3n_{\rm{H}}k_{\rm{B}}T$ with the ratio of specific heats $\gamma=5/3$. We adopt $Q=1.2 n^{2}_{\rm{H}}\Lambda \left(T\right)$ as the radiative loss function for optically thin emission \citep{colgan08}. Below 10,000 K, we set $\Lambda\left(T\right)$ to vanish because the plasma there is optically thick and no longer fully ionised. The term containing $\vec{\mathbf{\kappa}}=\kappa_{||}\mathbf{\hat{b}}\mathbf{\hat{b}}$ quantifies the anisotropic thermal conduction along the magnetic field lines with the Spitzer conductivity $\kappa_{||}$ as $10^{-6} \textit{T}$ $^{5/2}$ erg cm$^{-1}$ s$^{-1}$ K$^{-3.5}$. The flux of anisotropic thermal conduction has a ceiling, -sign$(\nabla T)5\phi \rho c^{3}_{s}$ as the saturated flux, and $c_{s}$ is the isothermal sound speed. The correction factor $\phi = 1$ is set according to the values suggested for the coronal plasma (\citealp{giuliani84}, \citealp{fadeyev02}, and references therein).

We employ a linear force-free magnetic field for the initial magnetic configuration, which is characterised by a constant cut-of-plane angle $\theta_0$ as follows:
\begin{displaymath}
 B_{x}=-B_{0} \cos \left( \frac{\pi x}{L_{0}} \right) \sin\theta_0 \exp\left(
 -\frac{\pi y \sin\theta_0}{L_{0}} \right)\,,
\end{displaymath}
\begin{displaymath}
 B_{y}=B_{0} \sin \left( \frac{\pi x}{L_{0}} \right) \exp\left(
 -\frac{\pi y \sin\theta_0}{L_{0}} \right)\,,
\end{displaymath}
\begin{equation}
 B_{z}=-B_{0} \cos \left( \frac{\pi x}{L_{0}} \right) \cos\theta_0 \exp\left(
 -\frac{\pi y \sin\theta_0}{L_{0}} \right)\,,
\label{bfield}
\end{equation}
with $\theta_0=30^\circ$, the angle between the arcade and the neutral line ($x = 0, y=0$). $L_{0}=80$ Mm is the horizontal size of our domain, and we adopt $B_{0}=50$ G.

For the initial thermal structure, we set a uniform temperature of 10,000 K below a height of 2.7 Mm and choose a temperature profile with height ensuring a constant vertical thermal conduction flux (i.e., $\kappa \partial{T}/\partial{y}$ = 2 $\times$ 10$^{5}$ erg cm$^{-2}$ s$^{-1}$) above this height as used in \citet{fang13} and \citet{xia12}. The initial density is then derived by assuming hydrostatic equilibrium with a number density of 1.2 $\times$ 10$^{15}$ cm$^{-3}$ at the bottom and the initial velocity field of all plasma is static. We employ a background heating rate decaying exponentially with height into the whole system all the time, $H_{0}=c_{0} \exp\left(-\frac{y }{\lambda_{0}} \right)$ where $c_{0}=10^{-4}$ erg cm$^{-3}$ s$^{-1}$ and $\lambda_{0}=50$ Mm. This heating is meant to balance the radiative losses and heat conduction related losses of the corona in its equilibrium state. With the above initial setup, the whole system now is out of thermal equilibrium. Therefore, we integrate the governing equations until the above configuration reaches a quasi-equilibrium state at 72 minutes after initialisation. Then we reset the time of the system back to zero for the next stage of simulation. As a result, the final relaxed state of the system is identified as the time when the maximal residual velocity in the simulation is less than 5 km s$^{-1}$. Panel (a), (b) and (c) in Fig.~\ref{f1} show the number density, temperature and AIA 131~\AA~ of the relaxed system, respectively. 

We use the MPI-parallelized Adaptive Mesh Refinement Versatile Advection Code {\it MPI-AMRVAC} \citep{amrvac,proth14,keppens14b} to run the simulation. An effective resolution of $1024 \times 640$ or an equivalent spatial resolution of 79 km in both directions is obtained through four AMR levels. Considering the left and right physical boundary, density, energy, $y$ and $z$ momentum components, $B_{y}$ and $B_{z}$ are set as symmetric, while $v_{x}$ and $B_{x}$ are taken antisymmetric to ensure zero face values. In the bottom boundary ghost cells, we use the primitive variables ($\rho, \textbf{v}, p, \textbf{B}$) to set all velocity components antisymmetric to enforce both no-flow-through (vertical) and no-slip (horizontal), while the $\mathbf{B}$ are fixed to the initial analytic expressions of equation~(\ref{bfield}), and the stratification of density is kept at pre-determined values from the initial condition, as well as the pressure. For the top conditions, we set all velocity components as antisymmetric, and adopt a discrete pressure-density extrapolation from the top layer pressure with a maximal temperature $T_{top} = 2 \times 10^{6}$ K. 

\subsection{Imaging and Spectroscopic Modelling}
To synthesize the observational features of SDO/AIA channels, we calculated the AIA temperature response function $K_{a}(n_{e},T)$[DN cm$^{5}$ s$^{-1}$]. The detail of the forward modelling method, can be found in \citet{yuan15}. The source of the forward modelling code (FoMo) is available at https://wiki.esat.kuleuven.be/FoMo. Then we assume that the flux $F_\alpha(x,y)$[ DN s$^{-1}$] will be integrated along the LOS for a width of W =1 Mm,
\begin{equation}
F_{\alpha}(x,y) = K_{\alpha}(n_{e},T)n^2_e\times\textrm{W}.
\end{equation}
We synthesized the AIA 94, 131~\AA~  channel emission which could image the flare loop at 6.4 MK and 10 MK, respectively, and the AIA 304~\AA~ channel emission line which would represent the transition region and top chromosphere (0.05 MK). 

Furthermore, we synthesized the extreme ultraviolet (EUV) emission intensity $I_{\lambda_{0}}$ [ergs cm$^{-2}$] of a specific spectral line $\lambda_{0}$ for optically thin plasma along LOS for a width of W =1 Mm. The details of the method can also be found in \citet{yuan15} and \citet{antolin13}, and the intensity is given by  
\begin{equation}
I_{\lambda_{0}}=\frac{A_b}{4\pi} G_{\lambda_{0}}(n_{e},T)n^2_e\times\textrm{W},
\end{equation}
where $A_b$ is the abundance of the emitting element relative to hydrogen, and $G_{\lambda_{0}}$ [ergs cm$^{-3}$ s$^{-1}$ ] is the contribution function that contains the terms relative to atomic physics, as a look-up table for SUMER Fe XIX 1118.1~\AA~ line in which most spectroscopic observations of standing slow waves were performed \citep{wang11}. 

\subsection{Triggering the Flare and Chromosphere Evaporation}
The flare is triggered by a finite duration heat pulse defined by the function $H_{1}$ located at the right footpoint between $x=22, 24$ Mm as the formula below (note that the heating rate $H$ in equation (3) is $H=H_{0}+H_{1}$). In our simulation, the heat pulse is controlled  by $f(t)$, which starts at time $t=0$ and switches off at $t=180$ seconds. The energy input by $H_{1}$ is around $3\times10^{28}$ erg s$^{-1}$, with an assumed thickness of 1 Mm along the third axis.  This energy input is suitable for a normal solar flare energy release. The energy of $H_{1}$ is quickly transported by thermal conduction to plasma at the footpoints.
\begin{equation}
\begin{array}{lrrr}
 H_{1}=c_{1} \exp(-(y-y_{c})^{2}/\lambda^2)f(t) & {\mathrm{ if }} & {A(x_{1},0)<A(x,y)<A(x_{2},0)}
 \end{array} 
\end{equation}
\begin{equation}
A(x,y)=\frac{B_{0}L_{0}}{\pi}\cos\left(\frac{\pi x}{L_{0}}\right)\exp\left(-\frac{\pi y \sin\theta_{0} }{L_{0}} \right) \,,
\end{equation}
\begin{equation}
f(t)=\left\{
\begin{array}{lrr}
t/30 & 0\leq t <30 &{\mathrm{s}}\\
1 & 30\leq t <150 &{\mathrm{s}}\\
(180-t)/30 & 150\leq t <180 &{\mathrm{s}}
\end{array} \right.
\end{equation}
where $c_{1}=16$ erg cm$^{-3}$ s$^{-1}$, $y_{c}=3$ Mm, $\lambda^{2}=10$ Mm$^{2}$, $x_{1}=24$ Mm, and $x_{2}=22$ Mm. The heat pulse is located close to the loop footpoint, i.e. $\approx$ 0.3 Mm above the transition region mimicking the footpoint heating by dissipated non-thermal particles. $A(x,y)$ is the magnetic potential depending on the location and decaying exponentially with height. Because the magnetic potential along a single magnetic field line is constant, we only add extra heating $H_{1}$ at one feet of a magnetic flux tube consisting of the magnetic field lines identified by $A(x,y)$ in the range of $x_{1}<x<x_{2}$. The large ratio between $c_{1}$ and $c_{0}$ as $1.6\times10^{5}$ highlights the extremely violent energy release of the solar flare.  

\section{Results and Discussion}\label{result}
\begin{figure}
 \centering
 \includegraphics[width=1.\textwidth]{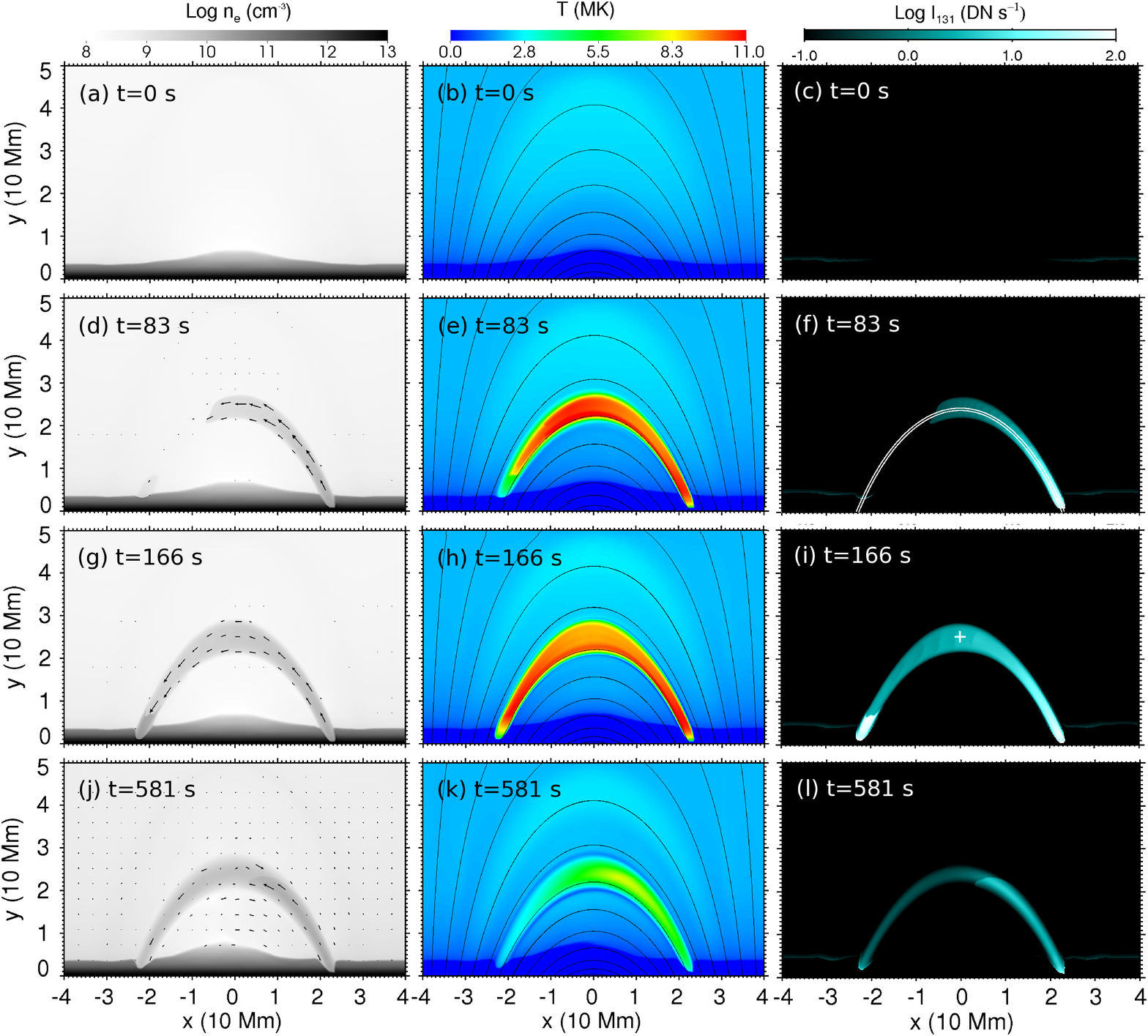}
 \caption{Temporal evolution of number density (left column), temperature (middle column), and synthesized AIA 131~\AA~emission (right column) images at $t\approx$ 0, 83, 166 and 581 seconds, respectively. The arrows in the left column mark the velocities and directions of the local plasma. The white lines in panel (f) denotes a fixed loop, defined by a field line with a fixed width of 1200 km. The cross in the panel (i) illustrates the initial location of the particle tracer ( $x=0$ Mm, $y=25$ Mm). An animation of this figure is available.}
 \label{f1}
\end{figure}

With the number density and temperature maps from the simulation, and the methods briefly presented in \textsection2.2, we can calculate the synthesized emission maps for both AIA and SUMER. We now discuss this simulation in more details. The three columns in Fig.~\ref{f1} display the temporal evolution of number density, temperature and synthesized AIA 131~\AA~emission intensity maps  at $t\approx$ 0, 83, 166 and 581 seconds, respectively. The extra flare heating $H_{1}$ at the right footpoint ($x=23$ Mm) imitates the explosive energy release with accumulated relativistic particles, suddenly dissipated in the upper chromosphere and transition region. This enormous energy heats the cold chromospheric plasma (around 0.02 MK) to an average temperature around 10 MK, as presented by the panel (e). The heated plasma is strongly evaporated into the confined loop as shown in panel (d).  The velocities of the plasma are represented by the arrows in panel (d), (g) and (j),  which show the directions of plasma movement. Compared with panel (d) and (e), panel (f) indicates that the synthesized emission in AIA 131~\AA~ channel is dominated by the density distribution rather than the temperature distribution. The front of the evaporated hot flow violently impacts the left footpoint ($x=-23$ Mm) at $t\approx$ 163 seconds with a speed up to 600 km/s. After the violent impact on the chromosphere, a reflected pattern rises up towards the loop apex as shown in panel (g), (h) and (i) at $t\approx$ 166 seconds. The AIA 131~\AA~ emission is mainly sensitive to 10 MK plasma. Still, there are subtle effects noted from comparing panel (h) and (i): the left footpoint of the loop has a strong synthesized AIA 131~\AA~emission in panel (i), but this would not be expected from the local temperature in panel (h). This strong emission is caused by the high compression and heating in the left footpoint, due to the wave impact. At $t\approx$ 446 seconds, the reflected pattern spreads back to the right footpoint ($x=23$ Mm). Then panel (j) at $t\approx$ 581 seconds represents the second reflection rising from the right footpoint and shows that the average density in the whole confined heated loop is clearly higher than at the beginning (panel (a)). Although the maximum of temperature in panel (k) decreases to around 8.5 MK, we still could observe the reflected pattern in AIA 131~\AA~ emission in panel (l). An animation of Fig.~\ref{f1} is available.

\begin{figure}
 \centering
 \includegraphics[width=\textwidth]{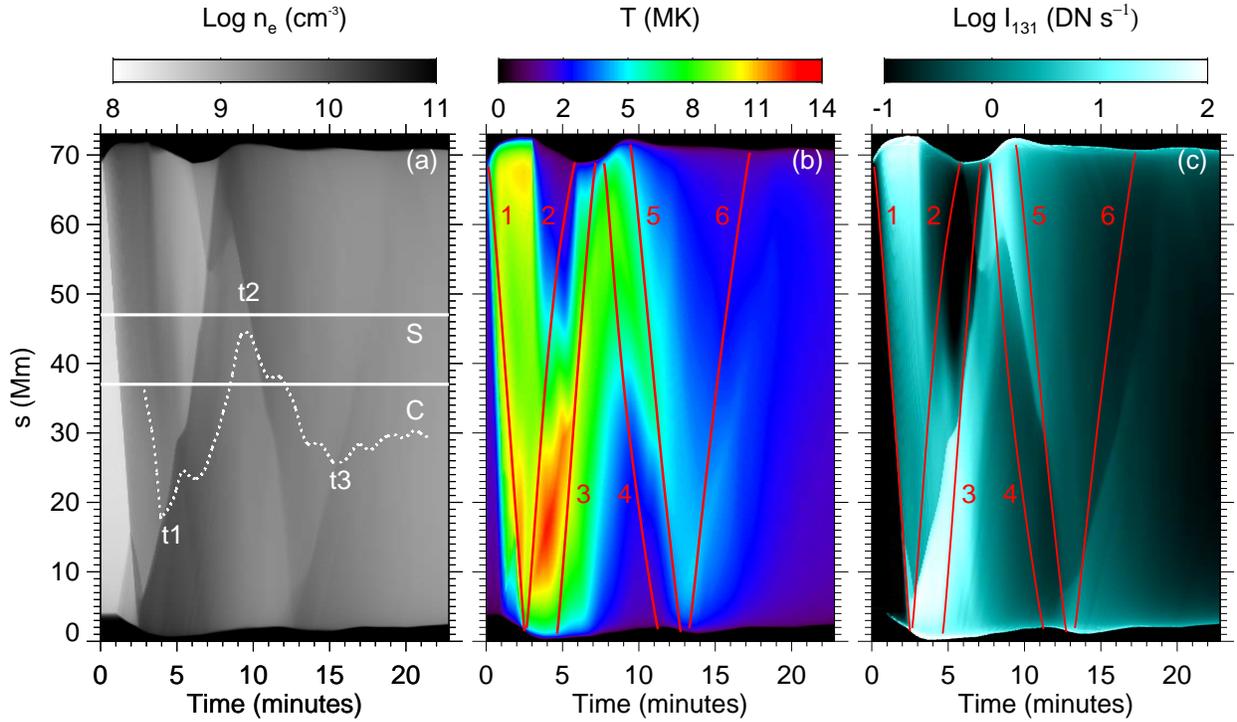}
 \caption{Values of number density, temperature and the synthesized AIA 131~\AA~ emission inside the loop shown as time-distance maps are displayed in panel (a), (b) and (c) , respectively. The six red solid lines in panel (b) and (c) show the paths of virtual particles propagating at the local sound speed. The dotted line in panel (a) shows the temporal evolution of the actual Lagrangian particle tracer. Position S and C are used to analyse the light curves in Fig.~\ref{f4} and \ref{f5}, respectively. t1, t2 and t3 indicate the times at which the tracer particle changes its direction. There is another online animation of evolution of virtual particles in panel (a) of Fig.~\ref{f2}.}
 \label{f2}
\end{figure}

In order to quantify and study these reflected patterns, we extracted a loop drawn by the white lines in panel (f) of Fig.~\ref{f1}. This fixed loop is defined by a field line identified at the start of the simulation, with a fixed width of 1200 km. We plot average values of the number density, temperature and synthesized AIA 131~\AA~ emission inside this loop to time-distance maps in panel (a), (b) and (c) of Fig.~\ref{f2}, respectively. The zero and end points along $s$ (vertical) axis in Fig.~\ref{f2} mean the left footpoint ($x=-23$ Mm) and right footpoint ($x=23$ Mm) in the domain, respectively. The reflected patterns are clearly seen as ridges in all three panels. We add a particle tracer at $t\approx 166$ seconds to trace and observe the movement of the plasma, especially when the reflected patterns sweep over it. The location of the particle tracer ($x=0,y=25$ Mm) is marked by a white cross in panel (i) of Fig.~\ref{f1}. The temporal evolution of the particle tracer movement is shown as the dotted line in panel (a) of Fig.~\ref{f2}. Because the third row of Fig.~\ref{f1} shows the moment when the front of the initial excited flow impacts the left footpoint ($s=2$ Mm) and the particle tracer is on the tail of the flow, the initial velocity of the particle tracer is towards the left footpoint ($s=2$ Mm). The particle moves downwards in panel (a) of Fig.~\ref{f2} until it is swept over by the reflected pattern at $t\approx 235$ seconds (this is t1), then it turns around towards the right footpoint ($s=70$ Mm). As shown in panel (a), the path of the particle displays three turnings (t1, t2 and t3), meaning that it is swept over by the reflected patterns three times. Although the particle is clearly in sync with the reflected patterns, the speed of the tracer is slower than the reflected patterns. The particle tracer behaves like a pendulum, where it is seen to oscillate back and forth, induced by the reflected patterns. There is another online animation of evolution of virtual particles in panel (a) of Fig.~\ref{f2}.

One important characteristic used by observational studies to identify this kind of solar propagating disturbances as a slow mode magnetosonic wave is the agreement between the estimated coronal sound speeds and speeds of the reflected propagating disturbances. Based on our simulation, we use a more accurate method to verify this agreement. The six red solid lines in panel (b) and (c) of Fig.~\ref{f2} show the paths of an imaginary particle propagating with the local sound speed, calculated by the local plasma temperature. We could consider these lines as paths of sound waves. In panel (c), line 1 represents the route of a sound wave initiated from the right footpoint ($s=70$ Mm) at the same time when the extra heating $H_{1}$ starts. We find that line 1 has a perfect agreement with the initial excited flow from the right footprint ($s=70$ Mm) to the left footpoint ($s=2$ Mm) in panel (c). This agreement indicates that the initial excited flow propagates with the sound speed. However, in panel (b) we find that the temperature rises earlier than the arrival of the flow (line 1). That is because of the faster propagation speed of the thermal conduction discontinuity, which also introduces a weak evaporation to slightly increase the density at the left footpoint ($s=2$ Mm) around $t\approx 1$ minutes as shown in panel (a). Line 2 represents the path of a sound wave propagating from the left footpoint ($s=2$ Mm) to the right footpoint ($s=70$ Mm),  assumed to set out when the excited flow propagating along line 1 impacts the left footpoint ($s=2$ Mm). Fig.~\ref{f2} show that there are AIA 131~\AA~ emission, density and temperature changes at the top end of line 2, the right footpoint ($s=70$ Mm). These increments indicate that a reflected wave propagates along line 2, and impacts to trigger another reflected wave rising from the right footpoint ($s=70$ Mm). The analysis of line 1 and line 2 confirm that these reflected patterns have a wave component.

However, unlike line 1, the synthesized AIA 131~\AA~ emission of the reflected pattern from the bottom end of line 1 does not behave similarly with line 2 in panel (c) of Fig.~\ref{f2}. The reflected pattern propagates slower than the sound wave along line 2, especially the part from the left footpoint ($s=2$ Mm) to $s=30$ Mm. Because the initially excited flow from the right footpoint ($s=70$ Mm) is triggered by a finite duration heating pulse, it relates to flows initiated within the same time range of $H_{1}$ from $t=0$ to $t=180$ seconds, which is clearly observed in all panels of Fig.~\ref{f2}. As a result, the rising reflected pattern at the left footpoint ($s=2$ Mm) from $t\approx161$ seconds encounters the rest part of the initially excited flow which still propagates towards the left footpoint ($s=2$ Mm). Panel (c) shows that this collision delays the propagation of the reflected pattern, indicating that both the excited flow and the reflected pattern contain a mass flow component. Line 3 in panel (c) is another path of a sound wave which arrives at the right footpoint ($s=70$ Mm) simultaneously with the main part of the reflected pattern. The middle piece of Line 3 shows that after passing through the ``collision" region, the reflected pattern propagates with the sound speed again. As well as for line 1, line 3 in the panel (b) temperature map shows that the thermal conduction discontinuity propagates faster than the sound wave. The behaviours of lines 1, 2 and 3 in panel (b) and (c) indicate these reflected patterns contain both wave and mass flow component. This is also confirmed by the line 4, 5 and 6. All of them tell the same story that the patterns observed in the synthesized AIA 131~\AA~channel emission is dominated by a repeatable wave component, and modulated by the mass flows where collisions can temporally retard or redirect actual mass flows. The wave-flow behaviour of the excited disturbance is reminiscent of shadow water waves \citep[e.g.][]{kundu04} for which a single pulse also gives a longitudinal displacement to trace particles. Another interesting phenomenon is that the highest temperature in our simulation is not produced initially by the flare heating $H_{1}$, but by the collisional compression of two patterns as shown in panel (b). 

\begin{figure}
 \centering
 \includegraphics[width=0.35\textwidth,angle=-90]{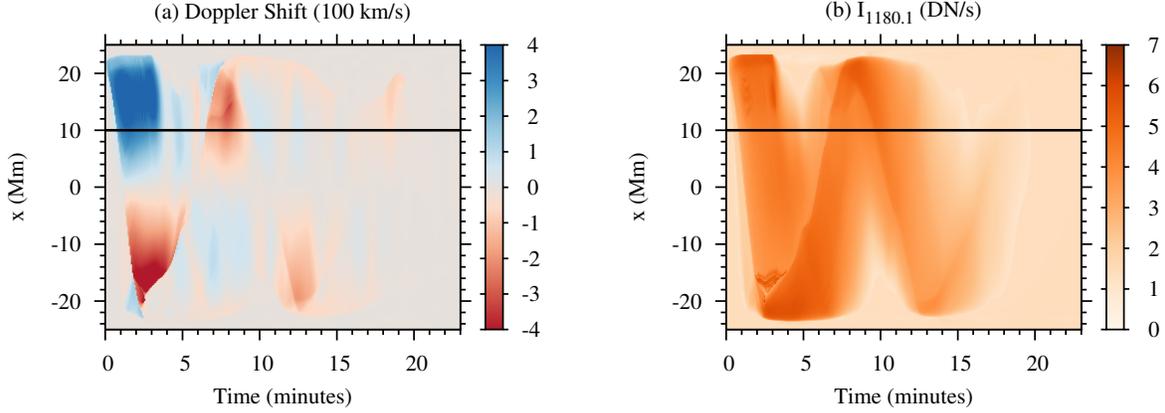}
 \caption{Doppler shift oscillations revealed by the synthesized
SUMER Fe XIX line emission maps: Panel (a) shows time series of Doppler shift in Fe XIX along a top view slit and panel (b) shows time series of the Fe XIX line intensity. The black full lines show the location of position S used in Fig.~\ref{f2} in this top slit view, and are used to analyse the light curves in Fig.~\ref{f4}. }
 \label{f3}
\end{figure}

\begin{figure}
 \centering
 \includegraphics[width=0.35\textwidth,angle=-90]{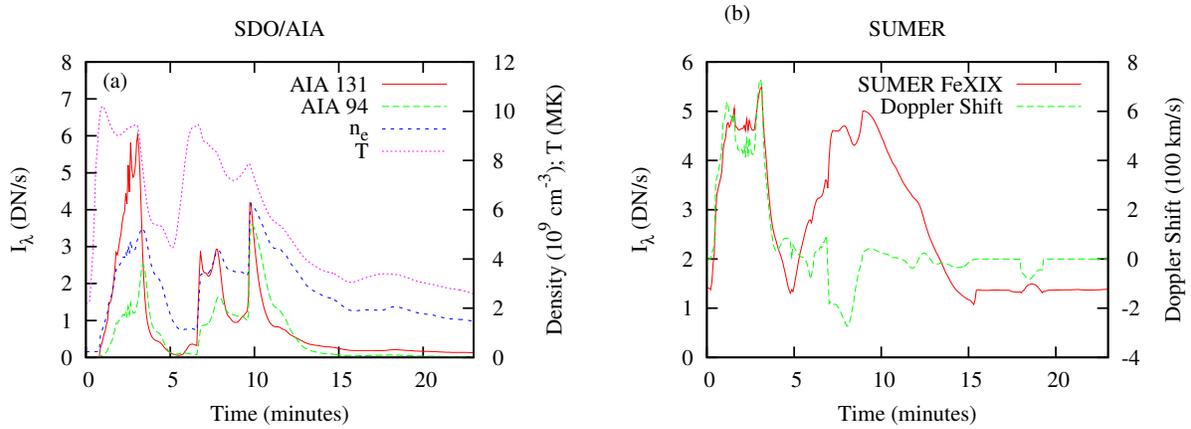}
 \caption{Panel (a) shows the temporal evolution of the synthesized AIA 131, 94~\AA~channel emission, number density and temperature for position S in Fig.~\ref{f2}; Panel (b) represents temporal evolution of the synthesized Doppler shift velocity and intensity of SUMER Fe XIX line for the black line in Fig.~\ref{f3}. }
 \label{f4}
\end{figure}

The slow magnetoacoustic waves observed by \citet{kumar15} are thought to be propagating waves, rather than standing waves as observed by \citet{wang11}. The standing waves have a unique characteristic, a quarter-period phase delay between the associated intensity variations and the Doppler signal. In Fig.~\ref{f3}, panel (a) shows a top slit view of the Doppler shift in the synthesized SUMER Fe XIX line, and panel (b) shows a top slit view of the intensity of the synthesized SUMER Fe XIX line emission. A top slit view means visual LOS from the top of the system ($y=50$ Mm) to the bottom of the system ($y=0$ Mm). The top view of the Doppler shift in panel (a) shows clear reflected patterns, and so does the intensity map in panel (b). We extract the black line ($x=10$ Mm) in Fig.~\ref{f3} to compare the synthesized Doppler signal and the associated intensity variations and identify which kind of wave we have in the simulation. Although the top slit views of the synthesized SUMER Fe XIX line emission integrate the quantities from the top to the bottom, the synthesized SUMER Fe XIX line only observes plasma with temperature greater than 6 MK, which only exists in the heated loop. So we extract position S in panel (a) of Fig.~\ref{f2} to compare with the black line in Fig.~\ref{f3},  which is located at the same position in the confined loop length. Panel (a) of Fig.~\ref{f4} shows the temporal evolution of the synthesized AIA 131, 94~\AA~ channel emission, number density and temperature of position S, and panel (b) represents the temporal evolution of the synthesized Doppler shift velocity and intensity of SUMER Fe XIX line of the black line, which is close to the right footpoint. So the second and third peaks of light curve in panel (a) don't mean peaks of the exactly second and third periods of waves, actually they indicate the time before and after the wave impact. The reason why the third peak is stronger than the second peak in AIA channels is that $I_{\lambda_{0}}$ of AIA channels are remarkably affected by the density. Since after the impact, the front of wave-flow reflects back and collides with the tail of itself and the density at position S increases as shown in panel (a) as well as the third peak of AIA channels. We find that  the number density, temperature and both AIA channels emission have an in-phase relationship as seen in panel (a), while the same is quantitatively  true for Doppler shift and intensity in SUMER Fe XIX line in panel (b). This suggests that these reflected patterns are propagating waves which show an in-phase relationship \citep{sakurai02}, rather than standing slow mode waves which show a quarter-period phase lag between velocity and intensity disturbances \citep{wang03a,yuan15}. The difference between AIA 131, 94~\AA~ channel emission is because AIA 131~\AA~ emission is more sensitive to higher temperatures around 10 MK, while AIA 94~\AA~ emission is sensitive to the temperature around 6.5 MK. This also explains the reason why AIA 94~\AA~ emission increases more than AIA 131~\AA~ emission at the third peak in panel (a) of Fig.~\ref{f3}, because at that moment the temperature decreases below 9 MK, while the density increases. 

\begin{figure}
 \centering
 \includegraphics[width=0.35\textwidth,angle=-90]{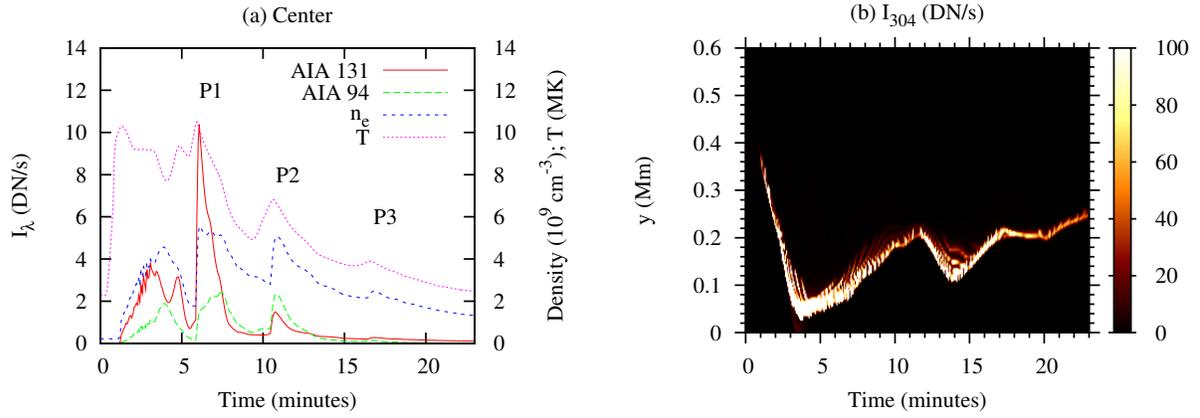}
 \caption{Panel (a) shows the temporal evolution of the synthesized AIA 131, 94~\AA~channel emission, number density and temperature for position C which locates at the loop apex as shown in Fig.~\ref{f2}; Panel (b) a Time-Distance plot illustrates the height and intensity variation of the plasma at the left footpoint (x= -23 Mm) imaged by the AIA 304 Channel.}
 \label{f5}
\end{figure}

In order to calculate the period of the reflected patterns, we extract another position C shown in panel (a) of Fig.~\ref{f2}. The position C is located at the apex of the loop, so the light curve of AIA 131, 94~\AA~ emissions of slice C in panel (a) of Fig.~\ref{f5} can reveal the more correct half period of the reflected patterns rather than panel (a) of Fig.~\ref{f4}. We identify three peaks of the reflected patterns based on the light curves of AIA 131, 94~\AA~ emission which all show strong damping afterwards. The time intervals between the three peaks are 285 and 360 seconds, so the total period would be 570 seconds and 720 seconds. Compared with the observational result in \citet{kumar15},  with 409 seconds for the period, the period in our simulation is longer. There are plenty of factors which could influence the period, such as the length and structure of the loop, the temperature inside the loop, etc. One possible reason for the longer period is the slower sound speed in our simulation, indicating that the reality should have even higher temperatures (greater than 10 MK). The period increase in our simulation is mainly affected by the large cooling, i.e the strong thermal conduction which is mainly responsible for the quick loop temperature decreases \citep{ofman02,wang03b} and little leakage of waves across the loop when the wave-flows impact the footpoint. When the temperature decreases during the propagation, and so does the sound speed, the period of the reflected patterns increases. The transition region at $x=-23$ Mm, shown by the AIA 304~\AA~ emission in panel (b) of Fig.~\ref{f5}, also displays an oscillation. This oscillation indicates the height variation of the transition region, due to the impact from reflected patterns. The height variation actually traces the energy and momentum exchange between coronal and chromospheric regions as leaked by the reflected patterns. \citet{yu2013} reports that quasi-periodic wiggles of microwave zebra pattern (ZP) structures can be associated with fast magnetoacoustic oscillations in a flaring active region. This oscillation of transition region associated with slow mode waves in our simulation may be observed in the future as well.

\section{Conclusion}
In this work, we performed a 2.5D MHD simulation to imitate the chromospheric evaporation and the following reflected patterns in a flare loop. We demonstrated that the periodic intensity variations captured by the synthesized AIA 131 and 94~\AA~emission images match well with previous observations \citep{kumar13,kumar15}. 

With a particle tracer, we confirmed that these reflected patterns contain a clear wave component, in their sound speed like propagation. Through predicted paths of sound waves, we also found that these reflected patterns are dominated by the wave component while modulated by mass flows. To sum up, the reflected patterns observed in our simulation contain both slow waves and mass flows. 

With the synthesized Doppler shift velocity and intensity maps in SUMER Fe XIX line emission, we confirmed that these reflected patterns are propagating slow mode waves rather than standing slow mode waves in our simulation, due to the in-phase relationship between Doppler shift and intensity.

From the light curves of the synthesized AIA 131, 94~\AA~ emission, we estimated the period of oscillations which increases from 570 seconds to 720 seconds during the observed three periods. The increase of the period was due to the decreasing loop temperature and sound speed, caused by the strong cooling. The height variation of the transition region shown in the synthesized AIA 304~\AA~ map may exhibit similar oscillations, correlated with the reflected patterns. This could be searched for in future observations.

\acknowledgments
The research has been sponsored by an Odysseus grant of
the FWO Vlaanderen. The results were obtained in the KU Leuven GOA project GOA/2015-014 and by the Interuniversity Attraction Poles Programme initiated by the Belgian Science Policy Office (IAP P7/08 CHARM). Part of the simulations used the infrastructure of the VSC - Flemish Supercomputer Center, funded by the Hercules Foundation and the Flemish Government - 
Department EWI.

\normalfont

\small
\begin{thebibliography}{}
\bibitem[Anfinogentov et al.(2013)]{anfinogentov13} Anfinogentov, S., Nakariakov, V.~M., Mathioudakis, M., Van Doorsselaere, T., \& Kowalski, A.~F.\ 2013, \apj, 773, 156 
\bibitem[Antolin \& Van Doorsselaere(2013)]{antolin13} Antolin, P., \& VanDoorsselaere, T.\ 2013, \aap, 555, A74 
\bibitem[Aschwanden et al.(2003)]{aschwanden03} Aschwanden, M.~J., 
Nightingale, R.~W., Andries, J., Goossens, M., 
\& Van Doorsselaere, T.\ 2003, \apj, 598, 1375 
\bibitem[Colgan et al.(2008)]{colgan08} Colgan, J., Abdallah, J., Jr., Sherrill, M.~E., et al.\ 2008, \apj, 689, 585
\bibitem[De Moortel \& Nakariakov(2012)]{moortel12} De Moortel, I., \& Nakariakov, V.~M.\ 2012, Royal Society of London Philosophical Transactions Series A, 370, 3193 
\bibitem[Fadeyev et al.(2002)]{fadeyev02} Fadeyev, Y.~A., Le Coroller, H., \& Gillet, D.\ 2002, \aap, 392, 735 
\bibitem[Fang et al.(2013)]{fang13} Fang, X., Xia, C., \& Keppens, R.\ 2013, \apjl, 771, L29 
\bibitem[Giuliani(1984)]{giuliani84} Giuliani, J.~L., Jr.\ 1984, \apj, 277, 605 
\bibitem[Goossens et al.(2002)]{goossens02} Goossens, M., Andries, J., \& Aschwanden, M.~J.\ 2002, \aap, 394, L39 
\bibitem[Gruszecki \& Nakariakov(2011)]{gruszecki11} Gruszecki, M., \& Nakariakov, V.~M.\ 2011, \aap, 536, A68 
\bibitem[Keppens et al.(2012)]{amrvac} Keppens, R., Meliani, Z., van Marle, A.~J., Delmont, P., Vlasis, A., \& van der Holst, B.\ 2012, JCP, 231, 718
\bibitem[Keppens \& Porth(2014)]{keppens14b} {Keppens}, R. \& {Porth}, O. 2014, Journal of Computational and Applied Mathematics, 266, 87
\bibitem[Kim et al.(2012)]{kim12} Kim, S., Nakariakov, V.~M., \& Shibasaki, K.\ 2012, \apjl, 756, L36 
\bibitem[Kumar et al.(2013)]{kumar13} Kumar, P., Innes, D.~E., \& Inhester, B.\ 2013, \apjl, 779, L7 
\bibitem[Kumar et al.(2015)]{kumar15} Kumar, P., Nakariakov, V.~M., \& Cho, K.-S.\ 2015, \apj, 804, 4 
\bibitem[Kundu et al.(2014)]{kundu04} Kundu, P.K., Cohen, I.M., \& Hu, H.H.\ 2004, Elsevier Academic Press
\bibitem[Liu \& Ofman(2014)]{liu14} Liu, W., \& Ofman, L.\ 2014, \solphys, 289, 3233 
\bibitem[Mariska(2006)]{mariska06} Mariska, J.~T.\ 2006, \apj, 639, 484 
\bibitem[Mariska(2005)]{mariska05} Mariska, J.~T.\ 2005, \apjl, 620, L67 
\bibitem[Nakariakov \& Ofman(2001)]{nakariakov01} Nakariakov, V.~M., \& Ofman, L.\ 2001, \aap, 372, L53 
\bibitem[Nakariakov et al.(2004)]{nakariakov04} Nakariakov, V.~M., Tsiklauri, D., Kelly, A., Arber, T.~D., \& Aschwanden, M.~J.\ 2004, \aap, 414, L25 
\bibitem[Nakariakov \& Melnikov(2009)]{nakariakov09} Nakariakov, V.~M., \& Melnikov, V.~F.\ 2009, \ssr, 149, 119 
\bibitem[Ofman et al.(2012)]{ofman12} Ofman, L., Wang, T.~J., \& Davila, J.~M.\ 2012, \apj, 754, 111 
\bibitem[Ofman \& Wang(2002)]{ofman02} Ofman, L., \& Wang, T.\ 2002, \apjl, 580, L85 
\bibitem[Porth et al.(2014)]{proth14} Porth, O., Xia, C., Hendrix, T., Moschou, S.~P., \& Keppens, R.\ 2014, \apjs, 214, 4
\bibitem[Roberts(2000)]{roberts00} Roberts, B.\ 2000, \solphys, 193, 139 
\bibitem[Sakurai et al.(2002)]{sakurai02} Sakurai, T., Ichimoto, K., Raju, K.~P., \& Singh, J.\ 2002, \solphys, 209, 265
\bibitem[Selwa et al.(2005)]{selwa05} Selwa, M., Murawski, K., \& Solanki, S.~K.\ 2005, \aap, 436, 701 
\bibitem[Selwa et al.(2007)]{selwa07} Selwa, M., Ofman, L., \& Murawski, K.\ 2007, \apjl, 668, L83 
\bibitem[Taroyan et al.(2005)]{taroyan05} Taroyan, Y., Erd{\'e}lyi, R., Doyle, J.~G., \& Bradshaw, S.~J.\ 2005, \aap, 438, 713 
\bibitem[Taroyan et al.(2007)]{taroyan07} Taroyan, Y., Erd{\'e}lyi, R., Wang, T.~J., \& Bradshaw, S.~J.\ 2007, \apjl, 659, L173 
\bibitem[Tsiklauri et al.(2004)]{tsiklauri04} Tsiklauri, D., Nakariakov, V.~M., Arber, T.~D., \& Aschwanden, M.~J.\ 2004, \aap, 422, 351 
\bibitem[Van Doorsselaere et al.(2011a)]{doorsselaere11a} Van Doorsselaere, T., De Groof, A., Zender, J., Berghmans, D., \& Goossens, M.\ 2011, \apj, 740, 90 
\bibitem[Van Doorsselaere et al.(2011b)]{doorsselaere11b} Van Doorsselaere, T., Wardle, N., Del Zanna, G., et al.\ 2011, \apjl, 727, L32 
\bibitem[Wang et al.(2002)]{wang02} Wang, T., Yan, Y., Wang, J., Kurokawa, H., \& Shibata, K.\ 2002, \apj, 572, 580 
\bibitem[Wang et al.(2003a)]{wang03a} Wang, T.~J., Solanki, S.~K., Curdt, W., et al.\ 2003, \aap, 406, 1105 
\bibitem[Wang et al.(2003b)]{wang03b} Wang, T.~J., Solanki, S.~K., Innes, D.~E., Curdt, W., \& Marsch, E.\ 2003, \aap, 402, L17 
\bibitem[Wang et al.(2007)]{wang07} Wang, T., Innes, D.~E., \& Qiu, J.\ 2007, \apj, 656, 598 
\bibitem[Wang(2011)]{wang11} Wang, T.\ 2011, \ssr, 158, 397 
\bibitem[Wang et al.(2013)]{wang13} Wang, T., Ofman, L., \& Davila, J.~M.\ 2013, \apjl, 775, L23 
\bibitem[Xia et al.(2012)]{xia12} Xia, C., Chen, P.~F., \& Keppens, R.\ 2012, \apj, 748, L26 
\bibitem[Yuan et al.(2014a)]{yuan14a} Yuan, D., Nakariakov, V.~M., Huang, Z., et al.\ 2014, \apj, 792, 41 
\bibitem[Yuan et al.(2014b)]{yuan14b} Yuan, D., Sych, R., Reznikova, V.~E., \& Nakariakov, V.~M.\ 2014, \aap, 561, A19 
\bibitem[Yuan et al.(2015)]{yuan15} Yuan, D., Van Doorsselaere, T., Banerjee, D., \& Antolin, P.\ 2015, \apj, 807, 98 
\bibitem[Yu et al.(2013)]{yu2013} Yu, S., Nakariakov, V.~M., Selzer, L.~A., Tan, B., \& Yan, Y.\ 2013, \apj, 777, 159 


\end{thebibliography}
\end{document}